# Relation between *ab initio* molecular dynamics and electron-phonon interaction formalisms

T.R.S. Prasanna

Department of Metallurgical Engineering and Materials Science

Indian Institute of Technology, Bombay

Mumbai, 400076 India

The relation between *ab initio* molecular dynamics formalism and the electron-phonon interaction formalism [P.B. Allen and V. Heine, J. Phys. C **9,** 2305 (1976)] is explored. The fundamental quantity obtained in the AIMD formalism – total energy for any configuration – is also obtained from the formalism (ES-DWF) that incorporates the role of Debye-Waller Factor in electronic structure calculations. The two formalisms are exactly equivalent and represent the direct and perturbation theory approaches to determine total energy. This equivalence allows either formalism to be used depending on the requirement – ES-DWF for *a priori* theoretical analysis and AIMD for *ab initio* modeling of the effect of thermal vibrations. Combining the two formalisms makes the ES-DWF formalism into an *ab initio* method and increases the range of problems that can be modeled *ab initio*. It is also theoretically possible to obtain self-consistent band structures from AIMD calculations. This study clarifies the incorrect assumptions regarding the two formalisms that exist in published literature. By combining the two formalisms and including self-energy effects, more accurate results can be obtained, *ab initio,* within the adiabatic approximation, than by using AIMD alone.

Thermal vibrations are universally present in all materials at finite temperatures and various formalisms have been developed to incorporate their role. One approach is the *ab initio molecular dynamics* (AIMD) formalism [1,2]. In this formalism, the electron distribution is assumed to be in equilibrium with every atomic configuration, which is the adiabatic or Born-Oppenheimer approximation. Wavefunctions are optimized and total energy calculated for any particular configuration. Forces are calculated (frequently from the Hellman-Feynman theorem) and the new atomic positions are obtained from Newton's laws. Wavefunctions are optimized and total energy calculated for the new atomic positions and this process is continued. In AIMD, electronic structure is solved for self-consistency at each time step. A variation of this method is the Car-Parinello MD formalism [3-5] where it is unnecessary to ensure self-consistency at each time step. AIMD formalism does not incorporate [1] self-energy effects that arise from electron-phonon interactions. The applications [1-5] of AIMD (or CPMD) are numerous. It gives information on equilibrium and non-equilibrium properties of materials. Ensemble averaging has to be performed to obtain equilibrium properties.

However, when only equilibrium properties are of interest, there exists another formalism that incorporates the role of thermal vibrations through electron-phonon interactions [6-8]. This formalism, which we will refer to as the EPI formalism, follows from the theory [6] developed to determine band structures at finite temperatures. This theory is based on the adiabatic approximation. In the EPI formalism, ensemble averaging over atomic displacements due to thermal vibrations is performed at the very beginning and the resulting electron energies are the ensemble averaged values. In this formalism, in addition to thermal expansion, electron-phonon interactions result in <u>two</u> terms that contribute to the electron energy, a) a Debye-Waller Factor component and b) self-energy component. The first term arises from the fact that the Fourier components of the core potential are altered at high temperatures by the DWF and can be represented as $V_G(T) = V_G(0)\ e^{-M(T)}$ where M is the DWF. That is, this term is the correction in electron energy due to the mean-square displacement of atoms or ions from their equilibrium positions caused by thermal vibrations. As this formalism has been developed using second order perturbation theory, where displacements are small, it cannot be used at high temperatures where the displacements are large [6]. The correct procedure to incorporate these effects is also described in the same article [6] as "*A higher order adiabatic perturbation summation can be accomplished by solving* $H_0 + \overline{H}_2 + \overline{H}_4 + ...$ *exactly (Keffer et. al. 1968) and then using the resulting temperature-dependent eigenfunctions and energies to calculate the self-energy terms*". That is, the first step is to incorporate the DWF in electronic structure calculations (which we refer to as ES-DWF formalism) and the next step is to use the results obtained to calculate the self-energy corrections. Therefore, symbolically the electron-phonon interaction formalism to incorporate the role of thermal vibrations can be represented as EPI = ES-DWF + SE, where SE contains the self-energy terms. Frequently, in practice, electronic structure calculations are performed by incorporating the DWF and the resulting band structure (that neglects self-energy contributions) is compared with experimental results. Because the role of thermal vibrations is incorporated through the DWF in the ES-DWF formalism, electronic structure has to be calculated <u>only once</u> at any temperature. The ES-DWF formalism is well established and is the first recourse [9-15] to explain the temperature dependence of valence electron properties in metals and semiconductors.

A clarification is necessary to avoid any possible confusion resulting from the slightly different definition of electron-phonon interactions in the EPI formalism of Ref.6-8. Conventionally, electron-phonon interaction is interpreted as resulting exclusively in self-energy effects. The role of thermal vibrations in altering total energy, density of states etc. that are studied by AIMD are not considered to be due to electron-phonon interactions. In the EPI formalism of Ref.6-8, all effects due to the role of thermal vibrations and not just self-energy effects are part of electron-phonon interactions. Hence, changes in electron energy due to different atomic configurations that result from thermal vibrations are also part of electron-phonon interactions.

From the above discussion, we see that when only equilibrium properties are of interest, two different formalisms exist that incorporate the role of thermal vibrations. *Hence, it is of great interest to explore the relationship between these two formalisms.* We show that the fundamental quantity obtained in the AIMD formalism – total energy for any configuration – can also be obtained from the ES-DWF formalism. The two formalisms are exactly equivalent and represent the direct and perturbation theory approaches to determine total energy. Using both formalisms allows their strengths to be combined and provides new physical insights in addition to extending the range of physical phenomena that can be modeled *ab initio*. This study also clarifies the incorrect assumptions regarding the two formalisms that exist in published literature. By combining the two formalisms and including self-energy effects, more accurate results can be obtained, *ab initio,* within the adiabatic approximation, than by using AIMD alone.

Firstly, both formalisms have been developed in the adiabatic approximation on general principles. Secondly, AIMD formalism incorporates the role of thermal vibrations except for self-energy effects[1]. The ES-DWF formalism is obtained by neglecting self-energy terms from the more general EPI formalism that incorporates the full effects of the role of thermal vibrations. In the AIMD formalism, ensemble averaging is performed at the end to obtain equilibrium properties. In the EPI formalism, ensemble averaging is performed during derivation itself and the results obtained are ensemble averaged quantities. The fundamental features of both the AIMD and ES-DWF formalisms are identical as they both incorporate the role of thermal vibrations (except for self-energy effects) within the adiabatic approximation. The order in which ensemble averaging is performed cannot affect the equilibrium values. Therefore, both AIMD and ES-DWF formalisms are theoretically exactly equivalent approaches to obtain equilibrium properties. This is established more rigorously below.

The fundamental quantity calculated in AIMD is $E_{AIMD}(\{u_l\})$, the total energy for a configuration $\{u_l\}$ of static lattice displacements. This total energy is obtained [1] directly by solving the electronic structure for a configuration $\{u_l\}$ of static lattice displacements.

We next derive the total energy in the ES-DWF formalism. Eq.3 of Ref.6 is the expression for electron energy, $E_{nk}(\{u_l\})$, for a configuration $\{u_l\}$ of static lattice displacements. After dropping the last term that gives the self-energy correction, Eq.3 of Ref.6 becomes

$$E_{nk}(\{u_l\}) = \varepsilon_{nk} + \langle nk|(H_1+H_2)|nk\rangle \quad \text{(AH3')}$$

where $H_1$ and $H_2$ are given by Eq.1 and Eq.2 of Ref.6 and $H_1$ accounts for anharmonicity. In principle, by summing $E_{nk}(\{u_l\})$ in Eq.AH3' over all $n$ and $k$, it is possible to obtain the total energy for the configuration $\{u_l\}$ of static lattice displacements. That is,

$$E_{DWF}(\{u_l\}) = E(\{0\}) + \sum_n \sum_k \langle nk|(H_1+H_2)|nk\rangle \quad (1)$$

where, $E(\{0\})$ is the total energy of the static lattice where all displacements are zero and is obtained from the first term in Eq. AH3' above.

Thus, there are two expressions, $E_{AIMD}(\{u_l\})$ and $E_{DWF}(\{u_l\})$, for the total energy for a given configuration $\{u_l\}$ of static lattice displacements. $E_{AIMD}(\{u_l\})$ is obtained in AIMD by directly solving the electronic structure for a configuration $\{u_l\}$ of static lattice displacements. $E_{DWF}(\{u_l\})$ is obtained from static lattice total energy by adding a perturbation correction. As long as the displacements are small so that perturbation theory is valid, clearly the two are equal and hence

$$E_{DWF}(\{u_l\}) = E_{AIMD}(\{u_l\}) \quad (2)$$

Eq.2 represents the equivalence of obtaining total energy directly and from perturbation theory. It follows that ensemble average total energies will also be equal, i.e.

$$\overline{E}_{DWF}(\{u_l\}) = \overline{E}_{AIMD}(\{u_l\}) \tag{3}$$

The above result is valid for small displacements where second order perturbation theory is valid, e.g. at low temperatures. When displacements are large, the total energy in the ES-DWF formalism is obtained [6] by solving the electronic structure by incorporating the role of DWF. As quoted earlier from Ref.6, this is equivalent to a higher order perturbation summation. Hence, the total energy obtained from such a calculation is the ensemble averaged total energy obtained in the perturbation theory framework, i.e. $\overline{E}_{DWF}(\{u_l\})$, but with no restriction that the displacements be small. In the AIMD formalism, $E_{AIMD}(\{u_l\})$ is obtained by taking ensemble average of total energy for each configuration that is obtained by directly from electronic structure calculations. Clearly, the results of direct calculations and perturbation theory are equivalent as long as perturbation theory is valid. Since $\overline{E}_{DWF}(\{u_l\})$ now represents the result of infinite order perturbation theory, Eq.3 is valid for all displacements.

The only reason for these two formalisms not being equivalent is if either formalism is theoretically deficient. Ref.8 shows that the adiabatic approximation, within which the EPI formalism [6] is derived, is valid for all materials at T > $\Theta_D$ (Debye temperature) and only fails for metals at low temperatures. The AIMD formalism is also developed within the adiabatic approximation as it assumes that the electrons are in equilibrium with the various atomic configurations that result from thermal vibrations. Therefore, the AIMD and ES-DWF formalisms are exactly equivalent at room and high temperatures for metals and at all temperatures for non-metals.

From the above discussion, it is clear that both AIMD and ES-DWF formalisms are exactly equivalent and represent the direct and perturbation theory approaches to determining total energies. This result is of great significance and the consequences are discussed below. Since both formalisms are complementary, their strengths and weaknesses are different and by using both formalisms the strengths of both formalisms can be combined as discussed below.

One of the main drawbacks of the ES-DWF formalism is that it is not an *ab initio* method and relies on experimentally determined lattice parameters (LP) and DWF. One of the main drawbacks of the AIMD formalism is that it is a computational technique with no possibility of *a priori* theoretical analysis or predictions of changes in properties due to thermal vibrations. Combining both formalisms allows these drawbacks to be overcome. The results of *ab initio* modeling from the AIMD formalism can now be interpreted in the theoretical framework of the ES-DWF formalism. For example, AIMD results on optical and dielectric properties can be combined with the theoretical analysis of Ref.16 for better understanding. Also, lattice parameters and mean (square) displacements (DWF) that can be obtained from AIMD formalism [3,17] can be used in the ES-DWF formalism. *Combining the two formalisms ensures that the ES-DWF formalism also becomes an ab initio method*.

The great advantage of this result is that *combining the AIMD and ES-DWF formalisms allows the ab initio determination of electronic band structures at finite temperatures.*

This is of great significance as in the ES-DWF formalism it is "*justified to speak of a Brillouin zone and a Fermi surface for other than T=0°K*" [12] and high temperature band structures are routinely displayed [12-15]. Therefore, combining the AIMD and ES-DWF formalisms implies that all properties, e.g. transport, optical and dielectric, that depend on the details of band structures, E(*k*), (discussed in Ref.6,7 of Ref.8) can also be modeled *ab initio*. Another example is that the changes in the band gap with temperature at high symmetry points in semiconductors [14,15], or even at any *k* point, can be modeled *ab initio*. Therefore, combining AIMD and ES-DWF formalisms vastly extends the range of problems that can be modeled *ab initio*.

In contrast, in the AIMD formalism, band structures are obtained from non self-consistent calculations. The authors of the popular software VASP state [18] that "*this is the only way to calculate the band structure, because for band-structure calculations the supplied k-points form usually no regular three-dimensional grid and therefore a self-consistent calculation gives pure nonsense!*". The equivalence between the AIMD and ES-DWF formalisms is independent of the computational technique employed for electronic structure calculations. Therefore, *in principle*, if the same computational technique is used in both AIMD and ES-DWF formalisms, the total energy obtained must be identical. This suggests that by using the same core potentials used in AIMD but corrected by DWF (that is obtained by AIMD), self-consistent band structures can be obtained that will *theoretically* have the same total energy as AIMD.

Another advantage of combining both formalisms is that the nuclear-nuclear repulsion energy, $E_{n-n}$, that must be evaluated to obtain total energy in either formalism can be easily obtained. It is difficult to evaluate changes in $E_{n-n}$ due to thermal vibrations in the AIMD formalism. In contrast, in the ES-DWF formalism, $E_{n-n}$ can be obtained at <u>*any temperature*</u> in a simple manner using Eq.5 of Ref.19. Using the same expression along with DWF obtained from AIMD, $E_{n-n}$ can be obtained very easily to be used with the AIMD formalism.

Because these formalisms are complementary, they give different physical insights. From Eq.3 it follows that the electronic DOS obtained from AIMD and ES-DWF formalism must be exactly equivalent. One interesting feature has been observed in the changes in DOS in both AIMD and ES-DWF formalisms. The authors of Ref.17 report that for Mo "*It is clear that the average DOS from the AIMD simulations is different from the DOS for the ideal lattice structures. One can clearly see that the thermal motion smears out most of the peculiarities of the DOS*". The authors have not provided any explanation for this observation. The ES-DWF formalism provides a ready explanation for this observation. Kasowski [11] explains that for Cd metal "*at higher temperatures the factor $e^{-W(k,T)}$ effectively reduces the potential and allows the density of states to become closer to the free-electron value*". Therefore, the most likely explanation for the AIMD observation [17] that peculiarities in DOS of bcc and fcc Mo are smeared at high temperatures is that it is due to the fact that the Fourier components of the potential, $V_G(T) = V_G(0)\ e^{-M(T)}$, decrease rapidly for large G due to the DWF. Hence, using both formalisms will lead to greater understanding and insights of any given problem.

The ES-DWF formalism is particularly useful in the study of alloy phase transitions [19]. Recent observations [20,21] of an isotope effect in magnetic phase transitions have been attributed to differences in exchange interactions due to different zero-point vibrations amplitudes. That is, the observed isotope effect <u>naturally suggests</u> that the ES-DWF formalism must be used to obtain a correct understanding of magnetic phase transitions. In addition, analysis within the ES-DWF formalism shows that the nuclear and core energy contributions to the alloy ordering energy are stored exclusively in superlattice lines [19] and this conclusion cannot be drawn from the AIMD formalism.

Also of great significance of the equivalence of the two formalisms is in understanding the high temperature thermodynamics of materials where the role of thermal vibrations is significant [22,23]. As discussed earlier, the EPI formalism is amenable to theoretical analysis and Ref.8 makes predictions on contributions of electron-phonon interactions to heat capacity, entropy etc. In contrast, AIMD formalism is not amenable to *a priori* theoretical analysis and no such predictions exist. However, AIMD allows *ab initio* modeling of changes due to thermal vibrations. Therefore, in published literature, experimental results are compared with the theoretical predictions of the EPI formalism and modeled in the AIMD formalism (see p-10, Ref. 22) thereby using both formalisms interchangeably. This interchangeable use is seen most clearly in Ref.23 whose author states "*Indeed, thermal disorder due to atomic vibrations broadens the EDOS, which in turn changes the phonon frequencies, and contributes to the vibrational entropy calculated from these frequencies. It can be shown that assigning this effect to the* *electronic or vibrational entropy is a matter of choice [18]; we choose to add this contribution to the electronic entropy since the average EDOS can be easily evaluated from AIMD simulations*" [23]. The Ref.18 of the author is the same as our Ref.8. Hence, the author has used the conclusions obtained from the EPI formalism [8] and implemented it using the AIMD formalism, thereby implicitly equating the two formalisms. Surprisingly, he has given neither any reference nor any justification for this assumed equivalence which is incorrect as discussed below.

The most significant result in Ref. 8 is the Brook's Theorem, Eq.6 of Ref.8, given by $\Delta E_{kn}(Q\mu) = \Delta \Omega_{Q\mu}(kn)$ from which other results are derived, including the one quoted above. But $\Delta E_{kn}(Q\mu)$ is given by Eq.2 of Ref.8 where the second term represents the self-energy contributions. Hence, all results of Ref.8 incorporate the role of self-energy contributions. It is well known [1] that AIMD formalism does not incorporate self-energy effects. Hence, it is clear that the equivalence implicitly assumed by the author of Ref.23 between AIMD and EPI formalisms is incorrect. The correct conclusion is that the AIMD and ES-DWF formalisms are equivalent as seen from the present study. While incorrect in detail and without any substantiation, the comment of the author [23] highlights the importance of establishing the equivalence between two formalisms. The equivalence between the AIMD and ES-DWF formalisms is established in the present study and hence, results from either formalism can be used interchangeably to understand the high temperature thermodynamics of materials.

Another insight that is obtained from the EPI formalism, but cannot be obtained from the AIMD formalism, is that the effect of the self-energy term on E($k$) is of the same order of magnitude [6] as the DWF term, within second order perturbation theory. Hence, self energy effects must be accounted for in order to obtain correct results. Their neglect implies that *ab initio* results of ES-DWF (or AIMD) calculations are approximate. This must be borne in mind when comparing results of *ab initio* calculations [17,22,23] with experimental data.

It is well known that AIMD formalism does not incorporate self-energy effects and it is necessary to go beyond the adiabatic approximation and adopt the time dependent Schrodinger's equation [1] or time dependent perturbation theory [8] to incorporate them. However, in the EPI formalism, it is possible to incorporate self-energy effects *within the adiabatic approximation* [6-8]. As quoted from Ref.6 earlier, for higher accuracy, band structures must be obtained in the ES-DWF formalism to which self-energy corrections must be added. By using the LP+DWF obtained from AIMD calculations in the ES-DWF formalism, band structure at high temperature can be obtained *ab initio*. To this result, if self-energy corrections are added, the resulting electron energies will be more accurate and obtained *within the adiabatic approximation*. Hence, we obtain an important result that by combining AIMD and ES-DWF formalisms and subsequently adding self-energy corrections, *it is theoretically possible to obtain more accurate results ab initio within the adiabatic approximation* than by using AIMD alone. This is the correct method to model *ab initio*, within the adiabatic approximation, all effects due to thermal vibrations.

In conclusion, the relation between two different formalisms that incorporate the role of thermal vibrations on ensemble average properties has been explored. The fundamental quantity obtained in the AIMD formalism – total energy for any configuration – can also be obtained from the formalism that incorporates the role of Debye-Waller Factor in electronic structure calculations. The two formalisms are exactly equivalent and represent the direct and perturbation theory approaches to determine total energy. The two formalisms are exactly equivalent and represent the direct and perturbation theory approaches to determine total energy. This equivalence allows either formalism to be used depending on the requirement – ES-DWF (or EPI) for *a priori* theoretical analysis and AIMD for *ab initio* modeling of the effect of thermal vibrations. Combining the two formalisms makes the ES-DWF formalism into an *ab initio* method and increases the range of problems that can be modeled *ab initio*. It is also theoretically possible to obtain self-consistent band structures from AIMD calculations. This study clarifies the incorrect assumption regarding the two formalisms that exist in published literature. By combining the two formalisms and including self-energy effects, more accurate results can be obtained, *ab initio,* within the adiabatic approximation, than by using AIMD alone.

1. R.M. Martin, *Electronic Structure*, Cambridge University Press, Cambridge (2004)
2. J. Hafner, J. Comput. Chem. **29**, 2044 (2008)
3. R. Car and M. Parinello, Phys. Rev. Lett. **55**, 2471 (1985)
4. G. Pastore, E. Smargiassi and F. Buda, Phys. Rev. A **44**, 6334 (1991)

5. M. C. Payne, M. P. Teter, D. C. Allan, T. A. Arias and J. D. Joannopoulous, Rev. Mod. Phys. **64**, 1046 (1992)

6. P.B. Allen and V. Heine, J. Phys. C **9,** 2305 (1976)

7. P. B. Allen, Phys. Rev. B **18**, 5217 (1978)

8. P.B. Allen and J. C. K. Hui, Z. Phys. B **37**, 33 (1980)

9. C. Keffer, T.M. Hayes and A. Bienenstock, Phys. Rev. Lett. **21,** 1676 (1968)

10. J. P. Walter, R.R.L. Zucca, M.L. Cohen and Y.R. Shen, Phys. Rev. Lett. **24,** 102 (1970)

11. R.V. Kasowski, Phys. Rev. **187**, 891 (1969)

12. R.V. Kasowski, Phys. Rev. B **8**, 1378 (1973)

13. T. V. Gorkavenko, S. M. Zubkova, and L. N. Rusina, Semiconductors, **41**, 661 (2007)

14. T. V. Gorkavenko, S. M. Zubkova, V. A. Makara and L. N. Rusina, Semiconductors, **41**, 886 (2007)

15. C. Sternemann, T. Buslaps, A. Shukla, P. Suortti, G. Doring and W. Schulke, Phys. Rev. B. **63** 094301 (2001)

16. B. Chakraborty and P.B. Allen, J. Phys. C **11,** L9 (1978)

17. C. Asker, A. B. Belonoshko, A. S. Mikhaylushkin and I. A. Abrikosov, Phys. Rev. B. **77** 220102(R) (2008)

18. G. Kresse, http://cms.mpi.univie.ac.at/vasp/vasp/node229.html

19. T. R. S. Prasanna, arXiv:0705.4382

20. P. A. Goddard, J. Singleton, C. Maitland, S. J. Blundell, T. Lancaster, P. J. Baker, R. D. McDonald, S. Cox, P. Sengupta, J. L. Manson, K. A. Funk, and J. A. Schlueter, Phys. Rev. B **78**, 052408 (2008)

21. H. Tsujii, Z. Honda, B. Andraka, K. Katsumata and Y. Takano, Phys. Rev. B **71**, 014426 (2005)

22. O. Delaire, M. Kresch, J. A. Munoz, M. S. Lucas, J. Y. Y. Lin and B. Fultz, Phys. Rev. B **77**, 214112 (2008)

23. V. Ozolins, Phys. Rev. Lett. **102** 065702 (2009)